\documentclass[pre,amsmath,amssymb]{revtex4}

\usepackage[dvips]{graphicx}
\usepackage{bm}

\begin{document}

\title{Synchrony of limit-cycle oscillators induced by random external
  impulses}
\author{Hiroya Nakao\footnote{Corresponding author :
    nakao@ton.scphys.kyoto-u.ac.jp}, Ken-suke Arai, Ken Nagai}
\affiliation{Department of Physics, Kyoto University, Kyoto 606-8502,
  Japan}
 \author{Yasuhiro Tsubo}
\affiliation{RIKEN Brain Science Institute, 2-1 Hirosawa, Wako,
  Saitama 351-0198, Japan}
\author{Yoshiki Kuramoto}
\affiliation{Department of Mathematics, Hokkaido University, Hokkaido
  060-0810, Japan}
\date{\today}

\begin{abstract}
  The mechanism of phase synchronization between uncoupled limit-cycle
  oscillators induced by common external impulsive forcing is
  analyzed.
  By reducing the dynamics of the oscillator to a random phase map, it
  is shown that phase synchronization generally occurs when the
  oscillator is driven by weak external impulses in the limit of large
  inter-impulse intervals.
  The case where the inter-impulse intervals are finite is also
  analyzed perturbatively for small impulse intensity.
  For weak Poissonian impulses, it is shown that the phase
  synchronization persists up to the first order approximation.
\end{abstract}

\maketitle

\section{Introduction}

When a limit-cycle oscillator is driven weakly by the same temporal
sequence of a fluctuating input, its phase tends to exhibit the same
dynamics repetitively among different realizations even if small
external disturbances distinguish each realization.
For example, a cortical neuron generates spikes more reproducibly when
it receives a fluctuating input current rather than a constant input
current~\cite{Mainen,Steveninck,Tsubo}.
This fluctuation-induced reproducibility of a single oscillator can be
interpreted as phase synchronization between uncoupled oscillators
driven by common external forcing, because repeated measurements on a
single oscillator using the same input is equivalent to a single
measurement on multiple identical oscillators.
It indicates the existence of a physical mechanism that statistically
stabilizes the limit-cycle orbit in the phase direction by a
fluctuating input.

There have been a variety of studies regarding this
phenomenon~\cite{Jensen,Kosmidis,Pakdaman,Gutkin,Ritt,Casado,Teramae,Nagai,Goldobin}.
Among them, Teramae and Tanaka~\cite{Teramae} made significant
progress in understanding its universality from the viewpoint of
nonlinear dynamics. Using the Stratonovich-Langevin equation resulting
from the phase reduction method~\cite{Winfree,Kuramoto,Pikovsky2},
they generally proved that limit-cycle oscillators always synchronize
in phase when they are driven by a vanishingly weak Gaussian-white
forcing (see also Goldobin and Pikovsky~\cite{Goldobin}).
Independently, we also analyzed this phenomenon in a different
setting~\cite{Nagai}. We assumed a simple random telegraphic forcing
to the oscillator that switches between two values randomly, which is
not necessarily vanishingly small. We reduced the dynamics of the
system to a pair of random maps, and generally showed that the
oscillators always synchronize in phase when the phase map is
monotonic.

In this paper, we consider yet another model of fluctuation-induced
phase synchronization. Specifically, we assume the external forcing to be
random impulses. Such a model has wide applications to various natural
phenomena, since impulsive noises are abundant in
nature~\cite{Snyder}.
For example, a cortical neuron interacts with other neurons via spike
trains, which are modeled as impulses in the simplest
approximation~\cite{Koch}.
Within this model, we can generally prove that the oscillators
actually undergo fluctuation-induced phase synchronization in the
limit of large inter-impulse intervals.
In addition, we can also discuss the case where the inter-impulse
interval is finite within this model.
Both of the previous analyses in Refs.~\cite{Teramae,Nagai} assumed
that the phase distribution of the oscillator is completely uniform on
the limit cycle, which corresponds to the assumption of vanishingly
weak or infinitely slow-switching external forcing.
However, in practical situations, such assumptions may not be valid,
and the phase distribution would generally be non-uniform. Thus, the
effect of non-uniform phase distribution on the phase synchronization
should be assessed.
By developing a perturbation theory for weak impulse intensity, we
discuss the effect of slight non-uniformity of the phase distribution
on the phase synchronization. Especially, we will show that the phase
synchronization persists even if the phase distribution becomes
slightly non-uniform for the Poissonian impulses.

Our analysis presented in this paper will extend the class of
fluctuation-driven limit-cycle oscillators that exhibit phase
synchronization, and will provide deeper insight into this phenomenon.

This paper is organized as follows. In section II, a general model of
impulse-driven limit-cycle oscillators is introduced, and phase
synchronization is demonstrated using two typical limit-cycle
oscillators. In section III, reduction of the dynamics of
impulse-driven oscillators to a random phase map is presented. In
section IV, stability in the phase direction is analyzed in the case
where the phase distribution is uniform. In section V, effect of
non-uniformity of the phase distribution is analyzed perturbatively
for small impulse intensity. Finally, Section VI gives the summary.

\begin{figure}[!htbp]
  \begin{center}
    \includegraphics[width=0.4\hsize,clip]{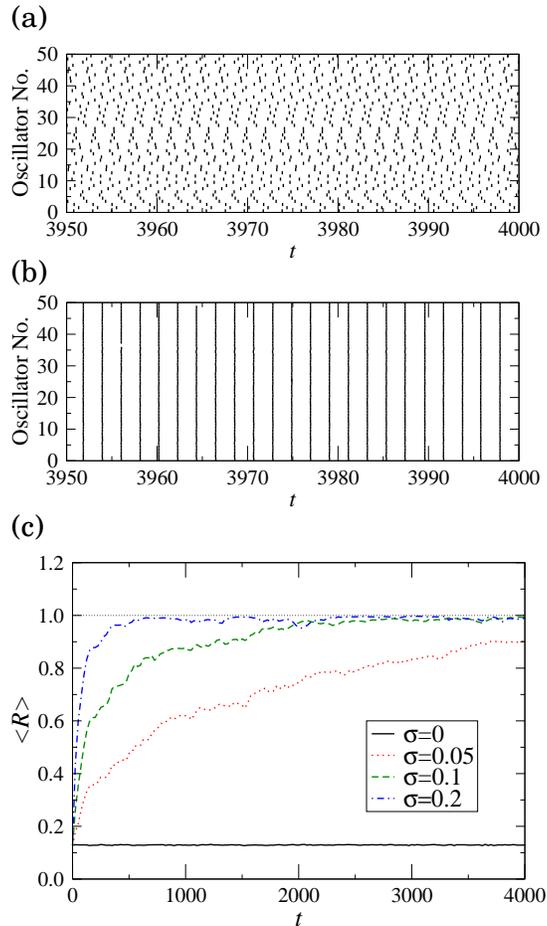}
    \caption{Synchronization of 50 Stuart-Landau oscillators driven by
      external impulses. (a) Zero-crossing events at $\sigma = 0$ and
      (b) at $\sigma = 0.1$. (c) Time sequence of the averaged modulus
      $\langle R \rangle$ of the order parameter calculated at $\sigma
      = 0, 0.05, 0.1$ and $0.2$.}
    \label{Fig:sl_raster_order}
  \end{center}
\end{figure}

\section{Phase synchronization induced by external impulses}

We first present a general model of limit-cycle oscillators driven by
a common external impulsive forcing, which we will analyze in later
sections. Then, before going into a general theory, we numerically
demonstrate phase synchronization induced by external impulses using
two typical models of limit-cycle oscillators, and briefly comment on
its mechanism.

\subsection{General Model}

We consider an ensemble of $N$ identical limit-cycle oscillators
subject to a common external impulsive forcing in the following
general form:
\begin{equation}
  {\dot {\bf X}_i}(t) = {\bf F}({\bf X}_i) + {\bf I}(t)
  \label{Eq:ode}
\end{equation}
for $i = 1, \cdots, N$, where ${\bf X}_i(t)$ represents the internal
state of the $i$-th oscillator at time $t$ and ${\bf F}$ its dynamics.
We assume that Eq.(\ref{Eq:ode}) has a single stable limit-cycle
solution ${\bf X}_0(t)$ in the absence of external impulsive forcing
${\bf I}(t)$, whose basin of attraction is the entire phase space
except some unstable fixed points.
External impulsive forcing ${\bf I}(t)$ is given by
\begin{equation}
  {\bf I}(t) = \sum_{n=1}^{\infty} {\bf e}_{n} \delta(t - t_{n}),
  \label{Eq:impulse}
\end{equation}
where $t_1, t_2, \cdots$ are occurrence times of the impulses, and
${\bf e}_{1}, {\bf e}_{2}, \cdots$ are random vectors representing the
``direction'' of the impulses.
When the oscillator receives an impulse at $t=t_n$, its state ${\bf
  X}_i$ is suddenly kicked by a random displacement ${\bf e}_n$ to a
new state ${\bf X}_i + {\bf e}_n$.

We assume that the direction of the impulse ${\bf e}$ is mutually
independent and identically distributed. We denote its probability
density by $Q({\bf e})$, which is normalized as $\int d{\bf e} Q({\bf
  e}) = 1$.
We also assume the interval $T$ between two successive impulses to be
independent and identically distributed. We denote its probability
density function by $P(T)$, which is normalized as $\int_0^{\infty} dT
P(T) = 1$.
We further assume the intervals to be sufficiently long, so that the
orbit kicked away from the limit cycle by an impulse can return to the
limit cycle before the next impulse. This is the necessary condition
for the phase description of the oscillator, which we adopt in the
following discussion. The time needed for this process of course
depends on the characteristic of the oscillator and on the intensity
of the impulses, which is very roughly of the order of the period of
the limit cycle.

In this paper, we mainly consider impulses generated by a Poissonian
process. Then Eqs.(\ref{Eq:ode}) and (\ref{Eq:impulse}) describe a
Poisson-driven Markov process~\cite{Snyder,Hanggi}.
In the Poissonian process, an impulse is generated with probability
$\nu$ in an infinitesimal time interval $\Delta t$. The probability
density $P(T)$ of the inter-impulse interval $T$ is given by the
exponential distribution
\begin{equation}
  P(T) = \frac{1}{\tau} \exp \left( -\frac{T}{\tau} \right),
  \label{Eq:expdist}
\end{equation}
where $\tau = \nu^{-1}$ determines the mean inter-impulse interval.
Of course, in this Poissonian case, there exists a certain probability
that the generated interval becomes shorter than the above-mentioned
return time of a kicked orbit to the limit cycle. In such a case, the
phase description fails to approximate the true dynamics
precisely. However, when $\tau$ is sufficiently large, such
probability becomes small, and the phase description would be a good
approximation statistically.

The probability density of the impulse direction $Q({\bf e})$ should
be chosen properly depending on the problem under consideration.  For
example, when we consider neural oscillators, usually only the
membrane potential can be stimulated experimentally by a current
injection. Therefore, the random vector ${\bf e}$ has only one
non-zero element corresponding to the voltage component of ${\bf X}$,
and we only need to consider one-dimensional probability density
$Q(\sigma)$, where $\sigma$ is the intensity of the current impulse.
In the following examples, we only treat the cases where the stimulus
can take a single fixed direction and intensity ${\bf e}_0$, namely
$Q({\bf e}) = \delta({\bf e} - {\bf e}_0)$, but we present our theory
in a general form so that it is also applicable to the case where the
stimulus takes various direction and intensity.

\subsection{Examples}

\subsubsection{Stuart-Landau oscillator}

Our first example is an ensemble of noisy Stuart-Landau
oscillators~\cite{Kuramoto} driven by a common sequence of Poissonian
random impulses with fixed intensity. The Stuart-Landau oscillator is
the simplest limit-cycle oscillator, which is derived as a normal form
of the super-critical Hopf bifurcation~\cite{Kuramoto}. The model is
described by
\begin{eqnarray}
  \dot{C_i}(t) &=& (1 + i c_0) C_i- (1 + i c_2) |C_i|^2 C_i + I(t) + \zeta_i(t)
  \label{Eq:SL}
\end{eqnarray}
for $i = 1, \cdots, N$, where $C_i$ is the complex amplitude of the
$i$-th oscillator, $c_0$ and $c_2$ are real parameters, $I(t)$
represents a sequence of external impulses, and $\zeta_i(t)$ is a
mutually independent complex Gaussian-white noise additionally
introduced to represent small external disturbances.
For simplicity, we drive only the real part of $C_i$ by the external
impulsive forcing $I(t)$, which is given by
\begin{equation}
  I(t) = \sigma \sum_{n=1}^{\infty} \delta(t - t_{n}),
\end{equation}
where the real parameter $\sigma$ represents the intensity of the
impulses, $t_{n}$ the time of occurrence of the impulse, and $\delta$
the Dirac's delta function.
The impulses are generated by a Poisson process with mean
inter-impulse interval $\tau$. When the oscillator receives an
impulse, its real component $\mbox{Re} C_i$ suddenly jumps by an
amount $\sigma$.
It is assumed that the complex Gaussian-white noise $\zeta_i(t)$ has
zero-mean, and whose variance is given by
\begin{eqnarray}
  \langle \mbox{Re} \zeta_i(t) \mbox{Re} \zeta_j(t') \rangle
  &=& D \delta_{i,j} \delta(t - t'), \cr
  \langle \mbox{Im} \zeta_i(t) \mbox{Im} \zeta_j(t') \rangle
  &=& D \delta_{i,j} \delta(t - t'), \cr
  \langle \mbox{Re} \zeta_i(t) \mbox{Im} \zeta_j(t') \rangle
  &=& 0,
\end{eqnarray}
where $D$ determines the noise intensity.
We fix the parameters at $c_0 = 2$, $c_2 = -1$, $\tau = 2$, and the
noise strength at $\sqrt{D} = 10^{-3}$.

We initially set the phase $\theta_i$ of each oscillator uniformly and 
randomly  on $[0,1]$ (see the next section for the precise definition
of the ``phase''), where the zero-crossing point of $C_i$ from
$\mbox{Im} C_i < 0 $ to $\mbox{Im} C_i > 0 $ is chosen as the origin
of phase, $\theta_i = 0$.
We then evolve the oscillators under the influence of the Poissonian
impulses and the weak Gaussian-white noise.
Figures~\ref{Fig:sl_raster_order}(a) and (b) plot zero-crossing events
of $C_i$ of $N=50$ Stuart-Landau oscillators by bars (so-called raster
plot) after transient for the cases $\sigma = 0$ and $\sigma = 0.1$.
It can be seen that the oscillators synchronize in phase when $\sigma
= 0.1$, whereas they do not synchronize at all when $\sigma = 0$.
To quantify the degree of synchronization, we introduce an order
parameter~\cite{Kuramoto}
\begin{equation}
  R \exp(2 \pi i \Theta) = \frac{1}{N} \sum_{i=1}^{N} \exp(2 \pi i \theta_i)
\end{equation}
using phase $\theta_i$ of each oscillator.
The modulus $R$ of this order parameter takes $R=1$ for complete
synchronization and $R=0$ for complete desynchronization.
Figure~\ref{Fig:sl_raster_order}(c) displays temporal evolution of the
modulus $\langle R \rangle$ averaged over 50 realizations from
different initial conditions. It gradually increases from a small
value to $1$ when $\sigma = 0.05$ and $\sigma = 0.1$, while it
constantly takes a small value when $\sigma = 0$.
Thus, the uncoupled Stuart-Landau oscillators driven by a common
sequence of Poissonian impulses synchronize in phase even if small
external disturbances exist.

\subsubsection{Hodgkin-Huxley model}

Our second example is an ensemble of the Hodgkin-Huxley neural
oscillators~\cite{Koch} driven by a common sequence of Poissonian
random impulses with fixed intensity.
It is given by the following set of equations~\cite{Koch}:
\begin{eqnarray}
  C_m \dot{V_i}(t) &=& G_{Na} m_i^3 h_i (E_{Na} - V_i) + G_{K} n_i^4 (E_{K} - V_i)
  + G_m (V_{rest} - V_i) + I_0 + I(t) + \xi_i(t), \cr
  \dot{m_i}(t) &=& \alpha_m (1 - m_i) - \beta_m m_i, \cr
  \dot{h_i}(t) &=& \alpha_h (1 - h_i) - \beta_h h_i, \cr
  \dot{n_i}(t) &=& \alpha_n (1 - n_i) - \beta_n n_i,
\end{eqnarray}
for $i=1, \cdots, N$, where $V_i$ represents the membrane potential of
the $i$-th neural oscillator, $m_i$ and $h_i$ the activation of its
sodium channel, and $n_i$ the activation of the potassium channel,
$I_0$ the constant input current, $I(t)$ the external impulsive
forcing, and $\xi_i(t)$ the additional external disturbances.
Parameters $G_{Na}$, $G_{K}$, and $G_{m}$ represent conductances of
the channels, $E_{Na}$, and $E_{K}$ represent their reversal
potentials, and $V_{rest}$ represents the rest voltage.
$\alpha_x$ and $\beta_x$ ($x = m, h, n$) are rate constants that are
given by the following equations:
\begin{eqnarray}
  \alpha_m &=& \frac{ 0.1 (25 - v) }{ \exp( \frac{ 25 - v }{10} ) - 1 }, \;\;\;
  \beta_m = 4 \exp( - \frac{v}{18} ), \cr
  \alpha_h &=& 0.07 \exp( - \frac{v}{20} ), \;\;\;
  \beta_h = \frac{1}{ \exp( \frac{30 - v}{10} + 1 ) }, \cr
  \alpha_n &=& \frac{ 0.01 (10 - v) }{ \exp( \frac{10 - v}{10} ) - 1 }, \;\;\;
  \beta_n = 0.125 \exp( - \frac{v}{80} ).
\end{eqnarray}
The parameters are fixed at the standard values presented in the
textbook~\cite{Koch}, i.e., $G_{Na} = 120$, $E_{Na} = 115$, $G_K =
36$, $E_K = -12$, $G_m = 0.3$, $V_{rest} = 10.613$ and $C_m = 1$.
We fix the constant input at $I_0 = 11$. When the external impulsive
forcing $I(t)$ is absent, this model exhibits stable limit-cycle
oscillation.
The external impulsive forcing is given by
\begin{equation}
  I(t) = \sigma \sum_{n=1}^{\infty} \delta(t - t_{n}),
\end{equation}
where $\sigma$ determines its intensity. The impulses are generated by
a Poissonian process with mean inter-impulse interval $\tau = 100$.
The external disturbance $\xi_i(t)$ is represented by a Gaussian-white
noise of mean $0$ and variance $D$. We fix the noise strength at
$\sqrt{D}=10^{-3}$ hereafter.
We define the zero-crossing event (``firing event'') of this
Hodgkin-Huxley neural oscillator as the moment at which the variable
$V_i$ changes its sign from $V_i < 0$ to $V_i > 0$.
We take this point as the origin, and define a phase along the limit
cycle~\cite{Winfree,Kuramoto}.

\begin{figure}[!htbp]
  \begin{center}
    \includegraphics[width=0.4\hsize,clip]{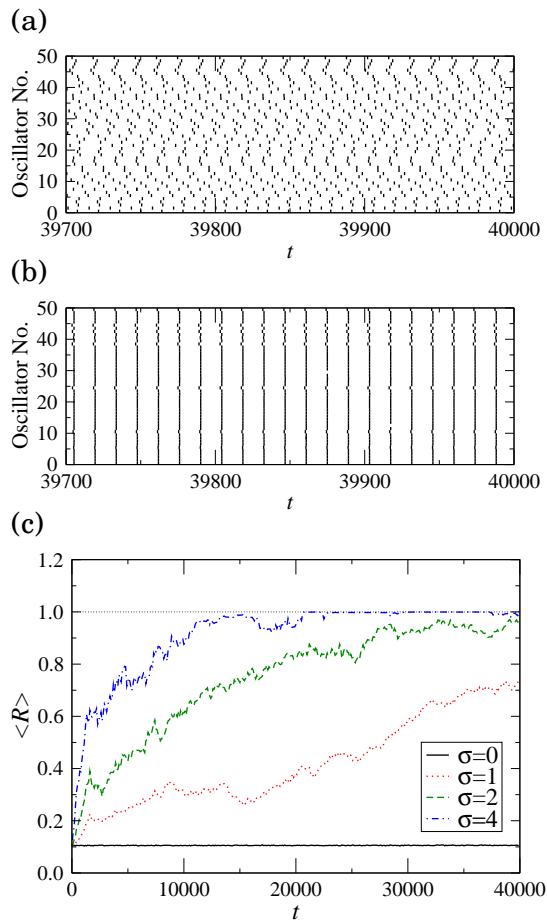}
    \caption{Synchronization of 50 Hodgkin-Huxley neural oscillators
      driven by external impulses. (a) Zero-crossing events at $\sigma
      = 0$ and (b) at $\sigma = 2$. (c) Time sequence of the averaged
      modulus $\langle R \rangle$ of the order parameter calculated at
      $\sigma = 0, 1, 2$ and $4$.}
    \label{Fig:hh_raster_order}
  \end{center}
\end{figure}

As in the previous case, we set the initial phases of the Hodgkin-Huxley
oscillators randomly, and evolve them under the effect of Poissonian
impulses and Gaussian-white noises.
Figures \ref{Fig:hh_raster_order}(a),(b) show zero-crossing events of
an ensemble of 50 Hodgkin-Huxley neural oscillators without or with
external impulses ($\sigma = 0$ or $\sigma = 2$) after the initial
transient.
The zero-crossing event well coincides with each other when $\sigma =
2$, i.e. the oscillators synchronize in phase after the initial
transient. Of course, they do not synchronize at all when $\sigma = 0$.
Figure \ref{Fig:hh_raster_order}(c) shows temporal evolution of the
modulus $\langle R \rangle$ of the order parameter averaged over 20
independent realizations at several values of the impulse intensity
$\sigma$.
When $\sigma$ takes $1$, $2$, or $4$, phase synchronization occurs and
$\langle R \rangle$ gradually increases from a small value to $1$.

\begin{figure}[!htbp]
  \begin{center}
    \includegraphics[width=0.4\hsize,clip]{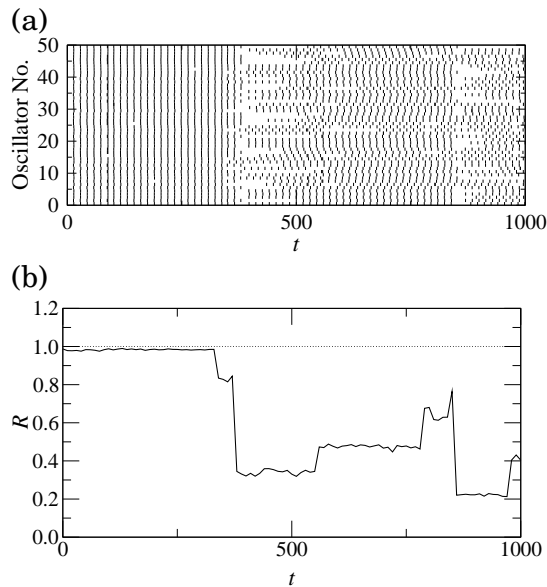}
    \caption{Desynchronization of 50 Hodgkin-Huxley neural oscillators
      induced by external impulses. (a) Zero-crossing events. (b)
      Temporal sequence of the modulus $R$ of the corresponding order
      parameter.}
    \label{Fig:hh_chaos}
  \end{center}
\end{figure}

It is also possible to observe impulse-induced desynchronization by
choosing the impulse intensity appropriately as shown in
Fig.~\ref{Fig:hh_chaos}(a), where the zero-crossing events of 50
Hodgkin-Huxley neural oscillators are plotted. In this case, the
external constant input is set at $I_0 = 9.8$, the intensity of the
external impulse at $\sigma = 6$, and the inter-impulse interval at
$\tau = 50$.
The phase of each oscillator is initially set at roughly the same
value, except tiny additional fluctuations of order $0.1$. To avoid
spurious complete synchronization due to numerical cutoff, small
external Gaussian-white noise with $\sqrt{D} = 10^{-3}$ is
additionally applied.
The phases of the oscillators scatter occasionally, and
correspondingly the order parameter drops as shown in
Fig.~\ref{Fig:hh_chaos}(b).
Thus, initial tiny phase differences among the oscillators can also be
enhanced by the external impulses.


\subsection{Some comments on the mechanism of impulse-induced phase
  synchronization}

The mechanism of phase synchronization induced by common external
input is basically a single-oscillator problem, though we consider an
ensemble of oscillators in the above examples.
The origin of the phase synchronization is the local stabilization of
each limit cycle in the phase direction due to the external impulses.
Namely, small phase disturbances of a single oscillator shrink
statistically, as we formulate in the following sections by reducing
the dynamics of each oscillator to a random phase map.
At the same time, it indicates the suppression of small difference in
phase between any pair of oscillators.
Due to random impulses, the phase of each oscillator diffuses on the
limit cycle in addition to the constant rotation.
Once two phases of any pair of the oscillators come close
accidentally, their difference can no longer grow but shrinks
statistically due to the local stability, leading to the synchrony of
the entire ensemble.
This mechanism has certain similarity to that of chaos synchronization
induced by common random forcing~\cite{Pecora,Pikovsky,Khoury,Toral}.

In the second example, we demonstrated that external impulses do not
only lead to phase synchronization but can also cause phase
desynchronization. Though we mainly focus on phase synchronization
in this paper, this fact is important in understanding that
fluctuation-induced phase synchronization is not a trivial phenomenon
but has some subtleties.

\section{Reduction to a random phase map}

In order to analyze the stability against phase disturbances, we first
reduce the dynamics of our impulse-driven limit-cycle oscillator to a
random phase map.

\subsection{Random phase map}

Following the standard procedure~\cite{Winfree,Kuramoto}, we define a
phase $\theta = \theta({\bf X}_0)$ along the limit cycle orbit ${\bf
  X}_0(t)$ in such a way that $\theta$ increases with a constant
angular velocity $\omega$. The phase $\theta$ is normalized by the
period of the limit cycle, so that its range is $[0,1]$ where $0$ and
$1$ represent the same phase.
This definition of phase can be extended to the entire phase space
except at phase-singular points, yielding a phase field $\theta({\bf
  X})$. It is achieved by identifying a point $P$ in the phase space
with a point $Q$ right on the limit cycle in such a way that the two
orbits started from $P$ and $Q$ asymptotically coincide. A set of
points that have equal phase is called an isochron. The entire phase
space is composed of isochrons with various phases.

In the absence of external impulses, the phase $\theta(t) =
\theta({\bf X}(t))$ obeys
\begin{equation}
  \dot{\theta}(t) = \omega
  \label{Eq:phasedyn}
\end{equation}
on the entire phase space (except at phase-singular points).
When the external impulses are given, the orbit is perturbed. Let us
assume that the orbit is on the limit-cycle at time $t = t_{n}-0$ ,
i.e. immediately before the $n$-th impulse (we say the orbit is ``on''
the limit cycle when it is sufficiently close to the limit cycle.)
We denote its location by ${\bf X}_{0}(t_{n}-0)$ and its phase by
$\theta_{n} = \theta(t_{n}-0)$.
We also denote the interval between the $n$-th impulse and the next
$(n+1)$-th impulse by $T_{n} = t_{n+1} - t_{n}$.
When the oscillator receives an impulse ${\bf e}_{n} \delta(t -
t_{n})$ at $t = t_{n}$, the orbit is kicked off the limit cycle and
jumps to a new phase-space point as
\begin{equation}
  {\bf X}(t_{n}+0) = {\bf X}_0(t_{n}-0) + {\bf e}_{n}.
\end{equation}
This new phase-space point ${\bf X}(t_{n}+0)$ is on a certain isochron
of the limit cycle, whose phase we denote by $\phi_{n}$ (unless it
is kicked exactly onto the phase-singular point, which rarely occurs).
We represent this mapping from $\theta_{n}$ to $\phi_{n}$ by
\begin{equation}
  \phi_{n} = F(\theta_{n}, {\bf e}_{n}) = \theta_{n} + G(\theta_{n}, {\bf e}_{n}),
  \label{Eq:phase_jump}
\end{equation}
which we call a ``phase map'' hereafter. In the second expression, we
split $F(\theta_{n}, {\bf e}_{n})$ into the trivial part $\theta_{n}$
that exists even without any impulses, and the non-trivial part
$G(\theta_{n}, {\bf e}_{n})$ arising from the impulse.
Since $F(\theta, {\bf e})$ is a phase map, it is a periodic function
on $[0,1]$. Therefore, $F(1, {\bf e}) = F(0, {\bf e})$ and $G(0, {\bf
  e}) = G(1, {\bf e})$ should hold (we should treat them in modulo
$1$).
As we discuss later, the above rule gives rise to an impulse-driven
phase equation of the Ito-type~\cite{Snyder,Hanggi}.

After the arrival of the $n$-th impulse, the oscillator evolves freely
with no external impulses from $t = t_{n}+0$ to $t = t_{n+1}-0$ for an
interval of $T_n = t_{n+1} - t_{n}$, and the phase changes from
$\phi_{n}$ to $\phi_{n} + \omega T_{n}$ during this interval. If $T_n$
is sufficiently large, the orbit evolves from ${\bf X}(t_{n}+0)$ to a
new point ${\bf X}_0(t_{n+1}-0)$ on the limit cycle.

Thus, corresponding to the evolution of the variable $ {\bf
  X}_0(t_{n}-0) \to {\bf X}(t_{n}+0) \to {\bf X}_0(t_{n+1}-0) $ from
$t = t_{n}-0$ to $t = t_{n+1}-0$, the phase evolves as $\theta_{n} \to
\phi_{n} \to \phi_{n}+\omega T_{n}$. Hence we obtain the following
evolution equation of the phase:
\begin{equation}
  \theta_{n+1}
  = \omega T_{n} + F(\theta_{n}, {\bf e}_{n})
  = \theta_{n} + \omega T_{n} + G(\theta_{n}, {\bf e}_{n}).
  \label{Eq:phasedyn_map}
\end{equation}
Since $T_{n}$ and ${\bf e}_{n}$ are random variables whose probability
density functions are given by $P(T)$ and $Q({\bf e})$ respectively,
this equation describes a random map.
When we consider Poissonian random impulses, the time step $n$ roughly
corresponds to the real time $t$ as $n \simeq t / \tau$, because the
mean inter-impulse interval is $\tau$.

If we go back to the continuous description, the dynamics of the phase
$\theta$ can be written as
\begin{equation}
  \dot{\theta}(t) = \omega +
  \sum_{n=1}^{\infty} G(\theta_{n}, {\bf e}_{n}) \delta(t - t_{n}).
  \label{Eq:phasedyn_ode}
\end{equation}
The external impulse is now explicitly multiplicative in this
equation. This impulse-driven phase equation is of Ito
type~\cite{Snyder,Hanggi}, namely, $G(\theta_n, {\bf e}_n)$ depends
only on the phase $\theta_n$ before the $n$-th impulse, which stems
from the rule we have assumed for the phase jump caused by an impulse.

\subsection{Relation to the phase response function}

According to the standard theory of phase
reduction~\cite{Winfree,Kuramoto}, when the orbit on the limit-cycle
${\bf X}_0$ at phase $\theta$ is kicked by a {\it weak} impulsive
force ${\bf p}$ to another isochron, its new phase $\phi$ is given by
a linear projection of the perturbation ${\bf p}$ on the gradient of
the phase field $\theta({\bf X})$ as
\begin{equation}
  \phi = \theta + {\bf Z}(\theta) \cdot {\bf p},
\end{equation}
where
\begin{equation}
  {\bf Z}(\theta) = \nabla_{\bf X} \theta({\bf X}) |_{{\bf X}_{0}(\theta)}  
\end{equation}
is the conventional phase response function representing the gradient
of $\theta({\bf X})$ on the limit cycle orbit ${\bf X}_0(t)$.
Comparing this equation with Eqs.(\ref{Eq:phase_jump}) and
(\ref{Eq:phasedyn_map}), we obtain
\begin{equation}
  F(\theta_{n}, {\bf e}_{n}) = \theta_{n} + {\bf Z}(\theta_{n}) \cdot {\bf e}_{n},
  \;\;\;
  G(\theta_{n}, {\bf e}_{n}) = {\bf Z}(\theta_{n}) \cdot {\bf e}_{n},
  \label{Eq:map_response}
\end{equation}
so that the phase dynamics can be described by
\begin{equation}
  \theta_{n+1} = \theta_{n} + \omega T_{n} + {\bf Z}(\theta_{n}) \cdot {\bf e}_{n}.
  \label{Eq:phasedyn_std_map}
\end{equation}
Thus, for sufficiently small ${\bf e}_{n}$, the phase map can simply
be represented using the inner product of the conventional phase
response function ${\bf Z}(\theta)$ and the direction of the impulse
${\bf e}_{n}$.

\subsection{Generalized Frobenius-Perron equation}

Temporal evolution of the probability density function (PDF)
$\rho(\theta, n)$ of the phase $\theta$ at time step $n$ is described
by a generalized Frobenius-Perron equation~\cite{Lasota}, which is
convoluted with a transition kernel $W(\theta)$ that represents random
shifting on the limit cycle for a random duration $T$ drawn from
$P(T)$, and is also averaged by the probability density $Q({\bf e})$
of impulse directions ${\bf e}$,
\begin{equation}
  \rho(\theta, n+1) =
  \int_0^1 d\phi W(\theta - \phi)
  \int d{\bf e} Q({\bf e})
  \int_0^1 d\psi \delta( \phi - F(\psi, {\bf e}) ) \rho(\psi, n),
  \label{Eq:pdfevol}
\end{equation}
where the argument $\theta - \phi$ of $W(\theta - \phi)$ should be
interpreted in modulo $1$.
In deriving this equation, we utilized the fact that $\theta_{n}$,
${\bf e}_{n}$, and $T_{n}$ are mutually independent ($\theta_{n}$
depends only on ${\bf e}_{1}, \cdots, {\bf e}_{n-1}$ and $T_{1},
\cdots, T_{n-1}$)~\cite{Lasota}.

For Poissonian impulses, the explicit form of the transition kernel
$W(\theta)$ can be calculated from Eq.(\ref{Eq:expdist}) by taking
into account the periodicity in $\theta$ and the Jacobian of the
transformation, which is given by
\begin{equation}
  W(\theta) = \frac{1}{\omega} \sum_{j=0}^{\infty} P(\frac{\theta + j}{\omega})
  = \frac{ e^{- \theta / ( \omega \tau ) } }
  { \omega \tau \left( 1 - e^{- 1 / \omega \tau} \right) }
  =
  \frac{A}{1 - e^{-A}} e^{-A \phi}
  \;\;\; (0 \leq \theta \leq 1),
  \label{Eq:transition_poisson}
\end{equation}
where we defined $A = 1 / \omega \tau$. Of course, it is normalized as
$\int_0^1 d\theta W(\theta) = 1$.

Sufficiently after the initial transient, the PDF $\rho(\theta, n)$ is
expected to reach a stationary state $\rho(\theta)$, but it is
generally difficult to calculate this stationary PDF analytically even
if the map $F(\theta, {\bf e})$ has a simple functional form.
In the following, we first analyze the limit of large inter-impulse
interval $\tau$ where the stationary PDF becomes uniform, and then
analyze the deviation of the stationary PDF from the uniform density
perturbatively for small impulse intensity.

\subsection{Examples of phase maps}

\subsubsection{Stuart-Landau oscillator}

\begin{figure}[!htbp]
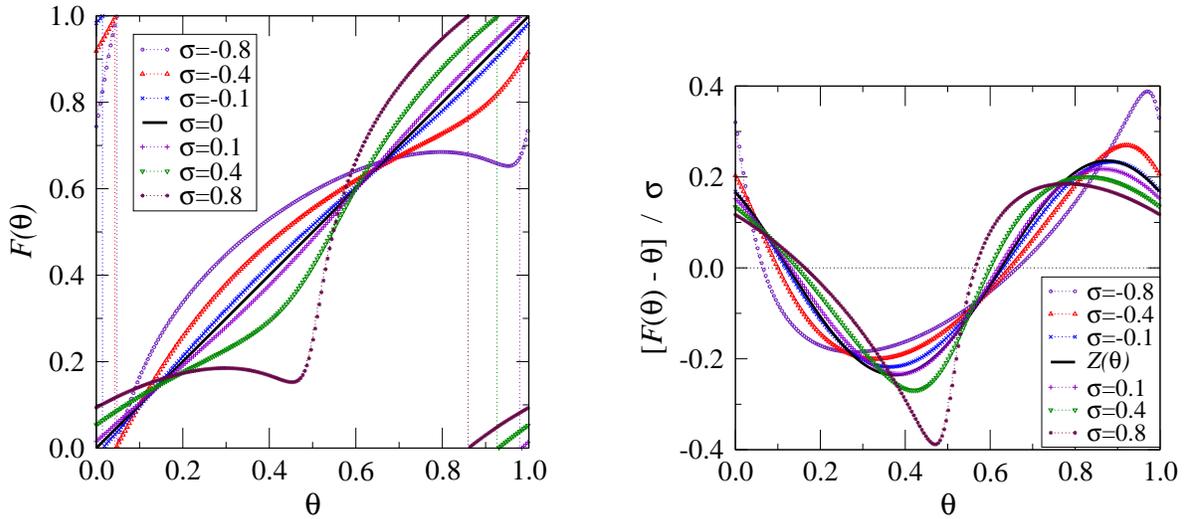

  \begin{center}
    \includegraphics[width=0.4\hsize,clip]{sl_F.eps}
    \hspace{1cm}
    \includegraphics[width=0.4\hsize,clip]{sl_G.eps}
    \caption{(a)Phase maps $F(\theta, \sigma)$ of the Stuart-Landau
      oscillator at several values of the impulse intensity $\sigma$.
      (b) Normalized phase response $\left[ F(\theta) - \theta \right]
      / \sigma$ at several values of the impulse intensity $\sigma$.
      Theoretical phase response function $Z_{\mbox{Re}}(\theta) =
      \sin(2 \pi \theta + 3 \pi / 4) / 3 \sqrt{2}$ that corresponds to
      the limit of $\sigma \to 0$ is also shown.  }
    \label{Fig:sl_phs_map}
  \end{center}
\end{figure}

Figure \ref{Fig:sl_phs_map}(a) plots numerically calculated phase maps
$F(\theta, \sigma)$ of the Stuart-Landau oscillator at several values
of the impulse intensity $\sigma$. As $|\sigma|$ becomes larger, the
phase map deforms from the trivial identity map noticeably, and
finally becomes non-monotonic when $\sigma \simeq \pm 0.8$.
Figure \ref{Fig:sl_phs_map}(b) displays the phase response normalized by
the impulse intensity $\left[ F(\theta, \sigma) - \theta \right] /
\sigma = G(\theta, \sigma) / \sigma$ at several values of $\sigma$.
If $\sigma$ is sufficiently small, Eq.(\ref{Eq:map_response}) should
hold, and the different curves corresponding to different values of
$\sigma$ should collapse.
The limiting curve at $\sigma \to 0$ gives the real component of the
phase response function ${\bf Z}(\theta)$. It can be analytically
calculated for the Stuart-Landau oscillator as $Z_{\mbox{Re}}(\theta)
= \sin(2 \pi \theta + 3 \pi / 4) / 3 \sqrt{2}$~\cite{Kuramoto}, which
is also shown in the figure.

\subsubsection{Hodgkin-Huxley model}

\begin{figure}[!htbp]
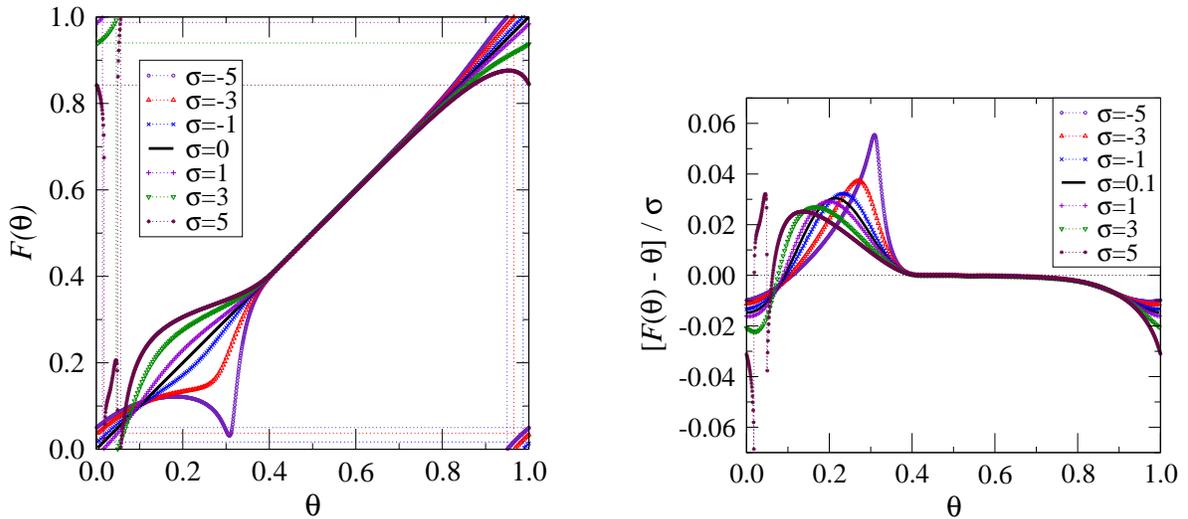

  \begin{center}
    \includegraphics[width=0.4\hsize,clip]{hh_F.eps}
    \hspace{1cm}
    \includegraphics[width=0.4\hsize,clip]{hh_G.eps}
    \caption{(a) Phase map $F(\theta, \sigma)$ of the Hodgkin-Huxley
      neural oscillator at several values of the impulse intensity
      $\sigma$.  (b) Normalized phase response $\left[ F(\theta) -
        \theta \right] / \sigma$ at several values of the impulse
      intensity $\sigma$.  The curve corresponding to $\sigma = 0.1$
      gives an approximation to the phase response function
      $Z_V(\theta)$ of the Hodgkin-Huxley neural oscillator.  }
    \label{Fig:hh_phs_map}
  \end{center}
\end{figure}

Figure \ref{Fig:hh_phs_map}(a) shows numerically calculated phase maps
$F(\theta, \sigma)$ of a Hodgkin-Huxley neural oscillator at several
values of the impulse intensity $\sigma$. As $|\sigma|$ becomes
larger, the phase map deforms from the trivial identity map and
finally becomes non-monotonic at $\sigma \simeq \pm 5$.
Figure \ref{Fig:hh_phs_map}(b) shows normalized phase response $\left[
  F(\theta, \sigma) - \theta \right] / \sigma  = G(\theta, \sigma) /
\sigma$ at several values of $\sigma$. It can be seen that the curves
actually collapse at small $\sigma$, and deviates at larger
$\sigma$. The liming curve at $\sigma \to 0$ gives the $V$-component
of the phase response function $Z_V(\theta)$. The curve corresponding
to $\sigma = 0.1$ in Fig.~\ref{Fig:hh_phs_map}(b) gives an
approximation to the phase response function.


\section{Stability in the phase direction}

Synchrony of uncoupled oscillators induced by external impulses is the
result of statistical stabilization of each oscillator against phase
disturbances. Such stability is characterized by the Lyapunov exponent
of the random phase map Eq.~(\ref{Eq:phasedyn_map}).

\subsection{Lyapunov exponent}

Let us consider the temporal evolution of a small deviation $\Delta
\theta_{n}$ from the original orbit $\theta_{n}$.
The linearized evolution equation of this small deviation is given by
\begin{equation}
  \Delta \theta_{n+1} = F'(\theta_{n}, {\bf e}_{n}) \Delta \theta_{n},
  \label{Eq:devel_deviation}
\end{equation}
where $F'(\theta_{n}, {\bf e}) = \left( dF(\theta, {\bf e}) / d\theta
\right)_{\theta = \theta_{n}}$.
At large time step $n$, $\Delta \theta_{n}$ expands as
\begin{eqnarray}
  \left| \frac{ \Delta \theta_{n} }{ \Delta \theta_{0} } \right|
  &=&
  \prod_{m=0}^{n-1} |F'(\theta_{m}, {\bf e}_{m})|
  \cr
  &=&
  \exp
  \left[
    \sum_{m=0}^{n-1} \log |F'(\theta_{m}, {\bf e}_{m})|
  \right]
  \cr \cr
  &\simeq&
  \exp \left( \lambda n \right),
\end{eqnarray}
where we introduced the Lyapunov exponent $\lambda$ of the map
$F(\theta, {\bf e})$ averaged over the PDFs $Q({\bf e})$ and
$\rho(\theta)$,
\begin{eqnarray}
  \lambda
  = \langle \langle \log |F'(\theta, {\bf e}) | \rangle \rangle_{\theta, {\bf e}}
  = \int_0^1 d\theta \rho(\theta) \int d{\bf e} Q({\bf e}) \log |F'(\theta, {\bf e})|.
\end{eqnarray}
If $\lambda$ is negative, $\Delta \theta_{n}$ shrinks on average, so
that the deviation from the original orbit is suppressed, whereas if
$\lambda$ is positive, small external disturbances will be enhanced.
Thus, the value of $\lambda$ gives a (local) condition for the phase
synchronization.

\subsection{Limit of large inter-impulse intervals}

As we mentioned previously, it is difficult to obtain the stationary
PDF $\rho(\theta)$ analytically.
However, when the inter-impulse interval $\tau$ is sufficiently large,
it can be approximated by a uniform density.
In the limit of large $\tau$, the transition probability tends to be
uniform, i.e. $W(\theta) \to 1$, which can easily be confirmed from
Eq.~(\ref{Eq:transition_poisson}) in the Poissonian case.
Correspondingly, the stationary phase PDF $\rho(\theta)$ approaches a
uniform density in the large $\tau$ limit:
\begin{equation}
  \rho(\theta) \to 1.
  \label{Eq:rhostationary}
\end{equation}

In this limit, we can obtain a sufficient condition of phase
synchronization for general limit-cycle oscillators: {\it when the
  phase map $F(\theta, {\bf e})$ is a monotonically increasing
  function of $\theta$, the Lyapunov exponent $\lambda$ is always
  non-positive}, namely, when $ F'(\theta, {\bf e}) = 1 + G'(\theta,
{\bf e}) > 0 $ is satisfied.
We can then bound $\lambda$ from above as
\begin{eqnarray}
  \lambda
  &=&
  \int_{0}^{1} d\theta \rho(\theta) \int d{\bf e} Q({\bf e})
  \log F'(\theta, {\bf e}) \cr \cr
  &\leq&
  \int_{0}^{1} d\theta \rho(\theta) 
  \log \left[
    \int d{\bf e} Q({\bf e}) F'(\theta, {\bf e})
  \right] \cr \cr
  &=&
  \int_{0}^{1} d\theta \rho(\theta)
  \log \left[ 1 + \int d{\bf e} Q({\bf e}) G'(\theta, {\bf e}) 
  \right] \cr \cr
  &\leq&
  \int_{0}^{1} d\theta \rho(\theta)
  \int d{\bf e} Q({\bf e}) G'(\theta, {\bf e}).
\end{eqnarray}
In the above transformation, we utilized Jensen's inequality $\int
d{\bf e} Q({\bf e}) \log[ g({\bf e}) ] \leq \log[ \int d{\bf e} Q({\bf
  e}) g({\bf e}) ]$ that holds for a concave function $\log[ \cdots ]
$, the normalized probability density $Q({\bf e})$, and a positive
scalar function $g({\bf e})$. The second inequality follows from
$\log(1+x) \leq x$.
By using the facts that $\rho(\theta) \equiv 1$ and $G(0, {\bf e}) =
G(1, {\bf e}) $, the upper bound of $\lambda$ can be calculated as
\begin{eqnarray}
  &&
  \int_{0}^{1} d\theta \rho(\theta)
  \int d{\bf e} Q({\bf e}) G'(\theta, {\bf e})
  \cr \cr
  &=&
  \int d{\bf e} Q({\bf e}) \int_{0}^{1} d\theta G'(\theta, {\bf e})
  \cr \cr
  &=&
  \int d{\bf e} Q({\bf e})
  \left[ G(1, {\bf e}) - G(0, {\bf e}) \right]
  =
  0.
\end{eqnarray}
Thus, for monotonically increasing $F(\theta, {\bf e})$, the Lyapunov
exponent $\lambda$ is always non-positive. The equality $\lambda = 0$
holds only when $F(\theta, {\bf e})$ is a trivial identity map for all
${\bf e}$, i.e.  $F(\theta, {\bf e}) = \theta$, which follows from the
equality condition of Jensen's inequality.
Therefore, small deviations from the original orbit always shrink by
applying random external impulses with large inter-impulse intervals,
when the phase map $F(\theta, {\bf e})$ is monotonic.

As we mentioned previously, when ${\bf e}$ is small, $G(\theta, {\bf
  e})$ can be represented using the phase response function ${\bf
  Z}(\theta)$ as $G(\theta, {\bf e}) \simeq {\bf Z}(\theta) \cdot {\bf
  e}$.  Since $F(\theta, {\bf e}) = \theta + G(\theta, {\bf e})$,
$F(\theta, {\bf e})$ is monotonically increasing with respect to
$\theta$ for sufficiently small ${\bf e}$.
Therefore, when the intensity of external impulses is small and the
mean interval between impulses is large, $\lambda$ always becomes
negative.

\subsection{Examples}

\subsubsection{Stuart-Landau oscillator}

As can be seen from Fig.~\ref{Fig:sl_phs_map}(a), the phase map
$F(\theta, \sigma)$ of the Stuart-Landau oscillator is monotonic as
long as $|\sigma|$ is small.
Therefore, the Stuart-Landau oscillator exhibits phase synchronization
induced by external impulses at such values of $|\sigma|$ for
sufficiently large inter-impulse intervals, as we demonstrated
previously.

\subsubsection{Hodgkin-Huxley model}

Similarly, as shown in Fig.~\ref{Fig:hh_phs_map}(a), numerically
calculated phase maps $F(\theta, \sigma)$ of the Hodgkin-Huxley neural
oscillator are monotonic when $|\sigma|$ is not so large.
Therefore, the Hodgkin-Huxley neural oscillators also exhibit
impulse-induced phase synchronization for such values of $|\sigma|$.
When $|\sigma|$ becomes large, the phase map can become quite complex,
which can lead to the impulse-induced phase desynchronization
mentioned previously.

\section{Effect of non-uniform phase distribution}

In the previous section, we discussed the limiting case of large
inter-impulse intervals, where the stationary PDF of the phase becomes
uniform. If the mean inter-impulse interval is not so large, the
stationary PDF would generally become non-uniform.
In this section, we first develop a perturbation theory to approximate
the non-uniform PDF for weak external impulses.
We then discuss the correction to the upper bound of the Lyapunov exponent
caused by the non-uniformity of the PDF.
In the following discussion, we assume that the intensity of external
impulses is sufficiently small, and that the phase map $F(\theta, {\bf
  e})$ is a strictly monotonically increasing function of $\theta$,
i.e. $F'(\theta, {\bf e}) > 0$.

\subsection{Perturbative solution to the generalized Frobenius-Perron
  equation}

As a first step, we calculate the deviation of the stationary PDF from
the uniform density perturbatively for small external impulses (up to
the second order). Our starting point is the generalized
Frobenius-Perron equation for the stationary PDF $\rho(\theta)$,
\begin{equation}
  \rho(\theta) = \int_0^1 d\phi W(\theta-\phi) \int d{\bf e} Q({\bf e})
  \int_0^1 d\psi \delta( \phi - F(\psi, {\bf e}) ) \rho(\psi).
  \label{Eq:FPS}
\end{equation}
We assume that the deviation of $F(\theta, {\bf e})$ from the identity
map $F(\theta, {\bf e}) = \theta$ is small,
\begin{equation}
  F(\theta, {\bf e}) = \theta + \epsilon G(\theta, {\bf e}),
\end{equation}
where we introduced a small parameter $\epsilon$ in order to control
the magnitude of the perturbation.
By using a well-known formula for the $\delta$-function,
Eq.(\ref{Eq:FPS}) can be rewritten as
\begin{equation}
  \rho(\theta) = \int_0^1 d\phi W(\theta-\phi) \int d{\bf e} Q({\bf e})
  \frac{\rho(\psi^{*}(\phi, {\bf e}))}{| F'(\psi^{*}(\phi, {\bf e}), {\bf e})|},
  \label{Eq:FPS2}
\end{equation}
where $\psi^{*}(\phi, {\bf e})$ is a solution to $F(\psi, {\bf e}) =
\phi$. We here used the fact that there exists only one solution,
because $F(\psi, {\bf e})$ is a monotonically increasing function of
$\psi$ (we do not consider the trivial case of $F(\psi, {\bf e})
\equiv \psi$, where $\rho(\theta)$ always becomes uniform).

We first calculate the solution $\psi^{*}(\phi, {\bf e})$ to $F(\psi,
{\bf e}) = \phi$ as a power series in $\epsilon$. We assume that the
solution can be expanded in terms of $\epsilon$ around the trivial
solution $\psi^{*}(\phi, {\bf e}) = \phi$ at $\epsilon = 0$ as
\begin{equation}
  \psi^{*}(\phi, {\bf e}) =
  \phi + \epsilon \psi_1^{*}(\phi, {\bf e}) + \epsilon^2 \psi_2^{*}(\phi, {\bf e}) + O(\epsilon^3).
\end{equation}
By inserting this expression to $F(\psi^{*}(\phi, {\bf e}), {\bf e}) =
\phi$, we obtain
\begin{eqnarray}
  \phi &=& \phi \;\;\; \mbox{at} \;\; O(\epsilon^0), \cr
  \psi_1^{*}(\phi, {\bf e}) + G(\phi, {\bf e}) &=& 0 \;\;\; \mbox{at} \;\; O(\epsilon^1), \cr
  \psi_2^{*}(\phi, {\bf e}) + G'(\phi, {\bf e}) \psi_1^{*}(\phi, {\bf e}) &=& 0 \;\;\; \mbox{at} \;\; O(\epsilon^2), \cr
  &\cdots&.
\end{eqnarray}
Thus, to the second order in $\epsilon$, the solution $\psi^{*}(\phi,
{\bf e})$ is approximated by
\begin{equation}
  \psi^{*}(\phi, {\bf e}) = 
  \phi - \epsilon G(\phi, {\bf e}) + \epsilon^2 G'(\phi, {\bf e}) G(\phi, {\bf e}) + O(\epsilon^3).
\end{equation}
Since $F'(\theta, {\bf e}) = 1 + \epsilon G'(\theta, {\bf e})$, it can
be expanded as
\begin{equation}
  F'(\psi^{*}(\phi, {\bf e}), {\bf e}) =
  1 + \epsilon G'(\phi, {\bf e}) - \epsilon^2 G''(\phi, {\bf e}) G(\phi, {\bf e}) + O(\epsilon^3).
\end{equation}
We also expand the stationary PDF $\rho(\theta)$ in a power series of
$\epsilon$ as
\begin{equation}
  \rho(\theta) = 1 + \epsilon \rho_1(\theta) + \epsilon^2 \rho_2(\theta) + O(\epsilon^3).
  \label{Eq:rho_expansion}
\end{equation}
Since $\rho(\theta)$ is normalized to $1$, $\int_0^{1} d\theta
\rho_1(\theta) = \int_0^{1} d\theta \rho_2(\theta) = 0$ should hold.
Inserting the above expansions into Eq.(\ref{Eq:FPS2}), we obtain
\begin{eqnarray}
  &&
  1 + \epsilon \rho_1(\theta) + \epsilon^2 \rho_2(\theta) + O(\epsilon^3)
  \cr \cr
  &=&
  \int_0^1 d\phi W(\theta-\phi) \int d{\bf e} Q({\bf e})
  \frac{
    1 + \epsilon \rho_1(\phi) + \epsilon^2 \left\{ \rho_2(\phi) - \rho_1'(\phi) G(\phi, {\bf e}) \right\} + O(\epsilon^3)
  }{
    1 + \epsilon G'(\phi, {\bf e}) - \epsilon^2 G''(\phi, {\bf e}) G(\phi, {\bf e}) + O(\epsilon^3)
  }
  \cr \cr
  &=&
  \int_0^1 d\phi W(\theta-\phi) \int d{\bf e} Q({\bf e})
  \left[
    1
    + \epsilon \left\{ \rho_1(\phi) - G'(\phi, {\bf e}) \right\}
    + \epsilon^2 \left\{ \rho_2(\phi) - H(\phi, {\bf e}) \right\}
    + O(\epsilon^3)
  \right]
\end{eqnarray}
where we utilized the fact that $ F'(\psi^{*}(\phi, {\bf e}), {\bf e})
> 0$, and defined
\begin{equation}
  H(\phi, {\bf e}) = 
  \rho_1(\phi) G(\phi, {\bf e}) - G'(\phi, {\bf e}) G(\phi, {\bf e}).
\end{equation}
Thus, the first order correction $\rho_1(\theta)$ to the uniform
stationary PDF $\rho_0(\theta) \equiv 1$ satisfies
\begin{eqnarray}
  \rho_1(\theta)
  &=&
  \int_0^{1} d\phi W(\theta-\phi) \int d{\bf e} Q({\bf e})
  \left\{  \rho_1(\phi) - G'(\phi, {\bf e}) \right\} \cr \cr
  &=&
  \int_0^{1} d\phi W(\theta-\phi)
  \left\{  \rho_1(\phi) - \langle G' \rangle_{\bf e}(\phi) \right\},
  \label{Eq:rho1}
\end{eqnarray}
and the second order correction $\rho_2(\theta)$ satisfies
\begin{eqnarray}
  \rho_2(\theta)
  &=&
  \int_0^{1} d\phi W(\theta-\phi) \int d{\bf e} Q({\bf e})
  \left\{  \rho_2(\phi) - H'(\phi, {\bf e}) \right\} \cr \cr
  &=&
  \int_0^{1} d\phi W(\theta-\phi)
  \left\{  \rho_2(\phi) - \langle H' \rangle_{\bf e}(\phi) \right\}.
  \label{Eq:rho2}
\end{eqnarray}
Here and hereafter,  for notational simplicity, we define an averaged
function $\langle f \rangle_{\bf e}(\theta)$ of a function $f(\theta,
{\bf e})$ over $Q({\bf e})$ as
\begin{equation}
  \langle f \rangle_{\bf e}(\theta) = \int d{\bf e} Q({\bf e}) f(\theta, {\bf e}),
\end{equation}
such as $\langle G' \rangle_{\bf e}(\phi) = \int d{\bf e} Q({\bf e})
G'(\phi, {\bf e})$ and $ \langle H' \rangle_{\bf e}(\phi) = \int d{\bf
  e} Q({\bf e}) H'(\phi, {\bf e})$.
We also define the Fourier transform between a function $f(\theta)$
and its coefficient $\tilde{f}(k)$ by
\begin{equation}
  f(\theta)
  =
  \sum_{k=-\infty}^{\infty} e^{2 \pi i k \theta} \tilde{f}(k),
  \;\;\;
  \tilde{f}(k)
  =
  \int_0^1 d\theta e^{- 2 \pi i k \theta} f(\theta).
\end{equation}
For example, the Fourier coefficients of $\rho_1(\theta)$ and
$G'(\theta, {\bf e})$ are denoted as $\tilde{\rho_1}(k)$ and
$\tilde{G'}(k, {\bf e})$, respectively.
The averaged Fourier coefficient $\langle \tilde{f} \rangle_{\bf
  e}(k)$ of $\tilde{f}(k, {\bf e})$ over $Q({\bf e})$ is similarly
defined as
\begin{equation}
  \langle \tilde{f} \rangle_{\bf e}(k) = \int d{\bf e} Q({\bf e}) \tilde{f}(k, {\bf e}),
\end{equation}
such as $\langle \tilde{G'} \rangle_{\bf e}(m) = \int d{\bf e} Q({\bf
  e}) \tilde{G'}(m, {\bf e})$ and $\langle \tilde{H'} \rangle_{\bf
  e}(m) = \int d{\bf e} Q({\bf e}) \tilde{H'}(m, {\bf e})$.

Equations (\ref{Eq:rho1}) and (\ref{Eq:rho2}) can be solved for
$\rho_1(\theta)$ and $\rho_2(\theta)$ through the Fourier transform,
which yields
\begin{eqnarray}
  \tilde{\rho_1}(m)
  &=&
  \int d{\bf e} Q({\bf e})
  \tilde{W}(m) \left\{ \tilde{\rho_1}(m) - \tilde{G'}(m, {\bf e}) \right\}
  =
  \tilde{W}(m) \left\{ \tilde{\rho_1}(m) - \langle \tilde{G'} \rangle_{\bf e}(m) \right\},
\end{eqnarray}
and
\begin{eqnarray}
  \tilde{\rho_2}(m)
  &=&
  \int d{\bf e} Q({\bf e})
  \tilde{W}(m) \left\{ \tilde{\rho_2}(m) - \tilde{H'}(m, {\bf e}) \right\}
  =
  \tilde{W}(m) \left\{ \tilde{\rho_2}(m) - \langle \tilde{H'} \rangle_{\bf e}(m) \right\}.
\end{eqnarray}
It can easily be shown that the equations at $m=0$ give trivial
relations, which should be neglected. We thus obtain
\begin{equation}
  \tilde{\rho_1}(m)
  =
  \frac{\tilde{W}(m)}{\tilde{W}(m) - 1}
  \langle \tilde{G'} \rangle_{\bf e}(m),
  \;\;\;
  \tilde{\rho_2}(m)
  =
  \frac{\tilde{W}(m)}{\tilde{W}(m) - 1}
  \langle \tilde{H'} \rangle_{\bf e}(m)
\end{equation}
for $m \neq 0$, and the corrections $\rho_1(\theta)$ and
$\rho_2(\theta)$ to the uniform density can be obtained as
\begin{equation}
  \rho_1(\theta)
  =
  \sum_{m \neq 0}
  \frac{\tilde{W}(m)}{\tilde{W}(m) - 1}
  \langle \tilde{G'} \rangle_{\bf e}(m)
  e^{2 \pi i m \theta},
  \;\;\;
  \rho_2(\theta)
  =
  \sum_{m \neq 0}
  \frac{\tilde{W}(m)}{\tilde{W}(m) - 1}
  \langle \tilde{H'} \rangle_{\bf e}(m)
  e^{2 \pi i m \theta}.
  \label{Eq:PDF_approx}
\end{equation}

For the Poissonian impulses, the transition probability $W(\theta)$ is
given by Eq.~(\ref{Eq:transition_poisson}), and its Fourier
coefficient $\tilde{W}(m)$ is given by
\begin{equation}
  \tilde{W}(m)
  = \int_0^1 d\theta W(\theta) e^{-2 \pi i m \theta}
  = \frac{A}{A + 2 \pi i m}
\end{equation}
for integer $m$. Therefore, the coefficient in
Eq.(\ref{Eq:PDF_approx}) is calculated as $ \tilde{W}(m) / [
\tilde{W}(m) - 1 ] = - A / (2 \pi i m)$.
Using this, we can calculate the first order correction to the phase
PDF as
\begin{eqnarray}
  \rho_1(\theta)
  &=&
  \sum_{m \neq 0} \frac{- A}{2 \pi i m}
  \langle \tilde{G'} \rangle_{\bf e}(m) \;
  e^{2 \pi i m \theta}
  =
  - A \sum_{m \neq 0} \langle \tilde{G} \rangle_{\bf e}(m) \;
  e^{2 \pi i m \theta} \cr \cr
  &=&
  - A \left[ \langle G \rangle_{\bf e}(\theta) - G_0 \right],
  \label{Eq:PDF_approx_poisson1}
\end{eqnarray}
where we defined $G_0 = \langle \tilde{G} \rangle_{\bf e}(0) =
\int_0^1 d\theta \langle G \rangle_{\bf e}(\theta)$, and utilized the
relation $\langle \tilde{G'} \rangle_{\bf e}(m) = (2 \pi i m) \langle
\tilde{G} \rangle_{\bf e}(m)$.
Similarly, the second order correction to the PDF can be calculated as
\begin{eqnarray}
  \rho_2(\theta)
  &=&
  \sum_{m \neq 0} \frac{- A}{2 \pi i m}
  \langle \tilde{H'} \rangle_{\bf e}(m) \;
  e^{2 \pi i m \theta}
  =
  - A \sum_{m \neq 0} \langle \tilde{H} \rangle_{\bf e}(m) \;
  e^{2 \pi i m \theta}
  \cr \cr
  &=&
  - A \left[ \langle H \rangle_{\bf e}(\theta) - \langle \tilde{H} \rangle_{\bf e}(0) \right],
  \cr \cr
  &=&
  - A \left[
    \langle \rho_1(\phi) G(\phi, {\bf e}) \rangle_{\bf e}
    -
    \langle G' G \rangle_{\bf e}(\theta) 
  \right]
  + const.
  \cr \cr
  &=&
  A^2
  \left[ \langle G \rangle_{\bf e}(\theta) \right]^2
  -
  A^2 G_0 \cdot
  \langle G \rangle_{\bf e}(\theta)
  +
  A \langle G' G \rangle_{\bf e}(\theta) 
  +
  const., 
  \label{Eq:PDF_approx_poisson2}
\end{eqnarray}
where we defined $\langle G' G \rangle_{\bf e}(\theta) = \langle
G'(\theta, {\bf e}) G(\theta, {\bf e}) \rangle_{\bf e}$. The constant
can be determined from the condition $\int_0^1 d\theta \rho_2(\theta)
= 0$.

Thus, in the Poissonian case, the averaged phase map $\langle G
\rangle_{\bf e}(\theta) $ directly appears at the first order
perturbation to the PDF, $\rho_1(\theta)$.
Since $A = 1 / \omega \tau$, the amplitude of the first order
perturbation scales as $\epsilon / \omega \tau$, namely, the ratio of
the impulse intensity to the non-dimensional time scale determined by
the period of the limit cycle and the inter-impulse intervals.
The second order perturbation $\rho_2(\theta)$ gives lowest-order
nonlinear contributions from the phase map.

\subsection{Upper bound of the Lyapunov exponent}

The Lyapunov exponent $\lambda$ is bounded from above as
\begin{eqnarray}
  \lambda
  &=&
  \langle \langle \log F'(\theta, {\bf e}) \rangle \rangle_{\theta, {\bf e}} \cr \cr
  &=&
  \int_0^1 d\theta \rho(\theta) \int d{\bf e} Q({\bf e}) \log F'(\theta, {\bf e}) \cr \cr
  &\leq&
  \int_0^1 d\theta \rho(\theta)
  \log \left\{ \int d{\bf e} Q({\bf e}) F'(\theta, {\bf e}) \right\}
  \cr \cr
  &=&
  \int_0^1 d\theta \rho(\theta)
  \log \left\{
    1 + \int d{\bf e} Q({\bf e}) \epsilon G'(\theta, {\bf e})
  \right\} \cr \cr
  &\leq&
  \int_0^1 d\theta \rho(\theta)
  \int d{\bf e} Q({\bf e}) \left\{ 
    \epsilon G'(\theta, {\bf e})
  \right\}.
\end{eqnarray}
Now, using Eq.~(\ref{Eq:rho_expansion}), correction to the upper bound
of the Lyapunov exponent can be expanded as
\begin{eqnarray}
  &&
  \int_0^1 d\theta \rho(\theta)
  \int d{\bf e} Q({\bf e}) \left\{ 
    \epsilon G'(\theta, {\bf e})
  \right\}
  \cr \cr
  &=&
  \int_0^1 d\theta
  \left\{
    1 + \epsilon \rho_1(\theta) + \epsilon \rho_2(\theta) + O(\epsilon^3)
  \right\}
  \int d{\bf e} Q({\bf e}) \left\{ 
    \epsilon G'(\theta, {\bf e})
  \right\}
  \cr \cr
  &=&
  \epsilon \int_0^1 d\theta \langle G' \rangle_{\bf e}(\theta)
  + \epsilon^2 \int_0^1 d\theta \rho_1(\theta) \langle G' \rangle_{\bf e}(\theta)
  + \epsilon^3 \int_0^1 d\theta \rho_2(\theta) \langle G' \rangle_{\bf e}(\theta)
  + O(\epsilon^4).
\end{eqnarray}
The first term corresponds to the (zero-th order) contribution from
the uniform component of the phase PDF, which vanishes (similarly to
the uniform case) irrespective of the functional form of the
transition kernel $W(\theta)$,
\begin{equation}
  \int_0^1 d\theta \langle G' \rangle_{\bf e}(\theta)
  = \langle G' \rangle_{\bf e}(1) - \langle G' \rangle_{\bf e}(0) = 0.
\end{equation}
For the Poissonian impulses, the first order correction to the upper
bound of $\lambda$ can be calculated using
Eq.(\ref{Eq:PDF_approx_poisson1}) as
\begin{eqnarray}
  &&
  \int_0^1 d\theta \rho_1(\theta) \langle G' \rangle_{\bf e}(\theta)
  \cr \cr
  &=&
  - A \int_0^1 d\theta \langle G \rangle_{\bf e} \langle G' \rangle_{\bf e}(\theta)
  + A G_0 \int_0^1 d\theta \langle G' \rangle_{\bf e}(\theta)
  \cr \cr
  &=&
  - A \left[ \frac{ \left\{ \langle G \rangle_{\bf e}(\theta) \right\}^2 }{ 2 } \right]_0^1
  + A G_0 \left[ \langle G \rangle_{\bf e}(\theta) \right]_0^1
  \cr \cr
  &=&
  0.
\end{eqnarray}
Thus, the upper bound of the Lyapunov exponent $\lambda$ is still zero
even if the first order correction to the uniform PDF is incorporated.
The second order correction to the upper bound can similarly be
calculated using Eq.(\ref{Eq:PDF_approx_poisson2}) as
\begin{eqnarray}
  &&
  \int_0^1 d\theta \rho_2(\theta) \langle G' \rangle_{\bf e}(\theta)
  \cr \cr
  &=&
  A^2 \int_0^1 d\theta
  \left\{ \langle G \rangle_{\bf e}(\theta) \right\}^2
  \langle G' \rangle_{\bf e}(\theta)
  -
  A^2 G_0 \int_0^1 d\theta
  \left\{ \langle G \rangle_{\bf e}(\theta) \right\}
  \langle G' \rangle_{\bf e}(\theta)
  \cr \cr
  &&+
  A \int_0^1 d\theta
  \langle G' G \rangle_{\bf e}(\theta)
  \langle G' \rangle_{\bf e}(\theta)
  +
  const. \int_0^1 d\theta
  \langle G' \rangle_{\bf e}(\theta)
  \cr \cr
  &=&
  \frac{A^2}{3} \left[ \left\{ \langle G' \rangle_{\bf e}(\theta) \right\}^3 \right]_0^1
  -
  \frac{A^2 G_0}{2} \left[ \left\{ \langle G' \rangle_{\bf e}(\theta) \right\}^2 \right]_0^1
  \cr \cr
  &&
  +
  const. \left[ \left\{ \langle G' \rangle_{\bf e}(\theta) \right\}^2 \right]_0^1
  +
  A \int_0^1 d\theta
  \langle G' G \rangle_{\bf e}(\theta) \langle G' \rangle_{\bf e}(\theta)
  \cr \cr
  &=&
  A \int_0^1 d\theta  \langle G' G \rangle_{\bf e}(\theta)
  \langle G' \rangle_{\bf e}(\theta).
\end{eqnarray}
Thus, only the term containing $\langle G' G \rangle_{\bf e}(\theta)$
in $\rho_2(\theta)$ gives non-vanishing contribution. Its sign cannot
be determined at this point unless the explicit functional form of
$G(\theta, {\bf e})$ is given. This term could make the upper bound of
the Lyapunov exponent slightly different from zero, but its effect is
only of the order of $\epsilon^3$.

Summarizing, the first order correction to the upper bound of the
Lyapunov exponent $\lambda$ by the non-uniformity of the phase PDF is
generally $O(\epsilon^2)$, but for the Poissonian impulses, it
vanishes.
Thus, the upper bound of $\lambda$ is still zero up to the first order
approximation.
The next order correction is only $O(\epsilon^3)$, which is quite
small when $\epsilon$ is small.
Therefore, the impulse-induced phase synchronization will, in most
cases, persist for weak Poissonian impulses even if the phase PDF
becomes slightly non-uniform for small $\epsilon$.

\subsection{Examples of Lyapunov exponents}

\subsubsection{Stuart-Landau oscillator}

\begin{figure}[!htbp]
  \begin{center}
    \includegraphics[width=0.4\hsize,clip]{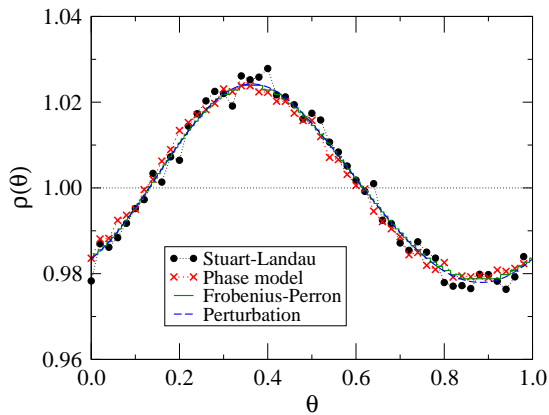}
    \caption{Stationary phase PDFs of the Stuart-Landau oscillator
      driven by external impulses obtained by (i) a direct simulation
      of the original model, (ii) a direct simulation of the
      corresponding phase model, (iii) numerically solving the
      corresponding Frobenius-Perron equation, and (iv) the
      perturbation theory.}
    \label{Fig:sl_pdf}
  \end{center}
\end{figure}

Stationary phase PDFs $\rho(\theta)$ of the Stuart-Landau oscillator
driven by external impulses at $\sigma=0.1$ and $\tau=2$ are shown in
Figure \ref{Fig:sl_pdf}.
The curves are obtained by (i) a direct simulation of the original
model Eq.(\ref{Eq:SL}) without Gaussian-white noise, (ii) a direct
simulation of the reduced phase model Eq.(\ref{Eq:phasedyn_ode}),
(iii) a numerical calculation of the stationary solution of the
corresponding Frobenius-Perron equation Eq.(\ref{Eq:FPS}), and (iv)
the perturbation theory up to the second order, respectively.
All curves agree well, which confirms the validity of our
approximation at least for small impulse intensity.
In this case, the first order perturbation already gives a nice fit to
the actual PDF, and the second order perturbation gives only a tiny
correction.

\begin{figure}[!htbp]
  \begin{center}
    \includegraphics[width=0.4\hsize,clip]{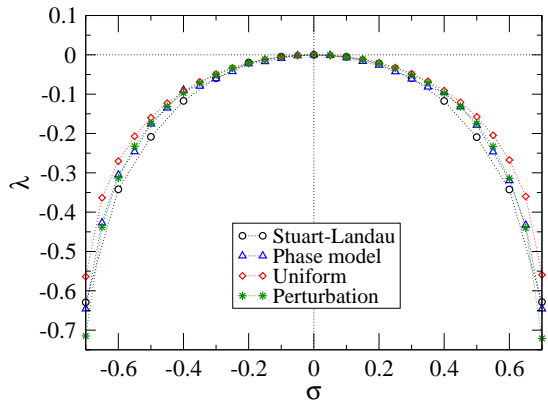}
    \caption{Dependence of the Lyapunov exponent $\lambda$ on the
      impulse intensity $\sigma$ obtained by (i) directly using the
      original model Eq.(\ref{Eq:SL}) without Gaussian-white noise,
      (ii) directly using the reduced phase model
      Eq.(\ref{Eq:phasedyn_ode}), (iii) a calculation using numerical
      phase maps assuming uniform phase PDF, and (iv) a calculation
      using numerical phase maps and the approximated phase PDF.  }
    \label{Fig:sl_lyap}
  \end{center}
\end{figure}

Figure \ref{Fig:sl_lyap} plots the Lyapunov exponent $\lambda$ as a
function of $\sigma$ at $\tau = 2$, which is obtained by (i) directly
using the original model Eq.(\ref{Eq:SL}) without Gaussian-white
noise, (ii) directly using the reduced phase model
Eq.(\ref{Eq:phasedyn_ode}), (iii) a calculation using numerical phase
maps and uniform phase PDF, and (iv) a calculation using numerical
phase maps and the approximated phase PDF.
Reflecting the symmetry of the limit cycle, the graph of
$\lambda$ is also symmetrical with respect to $\sigma=0$.
Since the phase map is always monotonically increasing in this range
of $\sigma$, the Lyapunov exponent $\lambda$ calculated assuming
uniform phase PDF (iii) is always non-positive and only becomes $0$
when $\sigma = 0$.
The Lyapunov exponent $\lambda$ calculated using approximate phase PDF
(iv) is also always non-positive. Since the correction to the uniform
PDF is small, the difference between (iii) and (iv) is also small.
Both curves agree well with the actual Lyapunov exponent obtained by
(i) and (ii).


\subsubsection{Hodgkin-Huxley model}

\begin{figure}[!htbp]
  \begin{center}
    \includegraphics[width=0.4\hsize,clip]{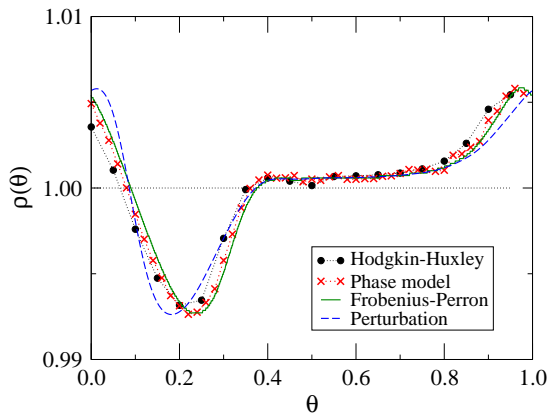}
    \caption{Stationary phase PDFs of the Hodgkin-Huxley neural
      oscillator driven by external impulses obtained by (i) direct
      simulation of the original model, (ii) direct simulation of the
      corresponding phase model, (iii) numerical solution of the
      corresponding Frobenius-Perron equation, and (iv) perturbation
      theory using Fourier coefficients numerically obtained from the
      phase map.}
    \label{Fig:hh_pdf}
  \end{center}
\end{figure}

Similarly, stationary PDFs of the Hodgkin-Huxley neural oscillator
driven by external impulses at $\sigma = 2$ and $\tau = 100$ are shown
in Figure \ref{Fig:hh_pdf}.
As previous, the curves represent the results obtained by (i) a direct
simulation of the original model, (ii) a direct simulation of the
reduced phase model, (iii) a numerical solution of the
Frobenius-Perron equation, and (iv) the perturbation theory.
Of course, the results of (ii) and (iii) give a nice fit to the actual
phase PDF obtained by (i). The result of perturbation theory (iv)
also gives a reasonable fit to the actual phase PDF.
As in the previous case, the first order perturbation gives a good fit to
the actual PDF, and the second order perturbation gives only a tiny
correction.

\begin{figure}[!htbp]
  \begin{center}
    \includegraphics[width=0.4\hsize,clip]{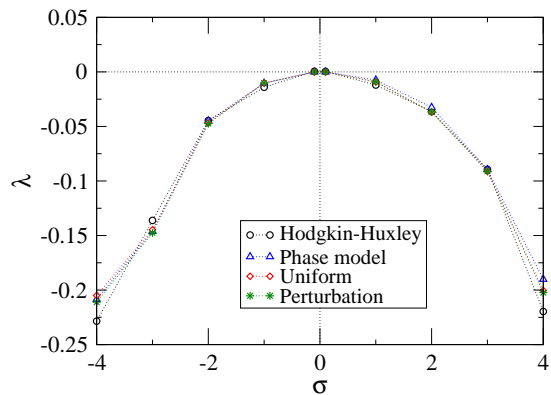}
    \caption{Lyapunov exponent $\lambda$ obtained as a function of
      the impulse intensity $\sigma$. by (i) directly using the
      original model, (ii) directly using the reduced phase model,
      (iii) a calculation using numerical phase maps assuming uniform
      phase PDF, and (iv) a calculation using numerical phase maps and
      the approximated phase PDF.}
    \label{Fig:hh_lyap}
  \end{center}
\end{figure}

Figure~\ref{Fig:hh_lyap} plots the Lyapunov exponent $\lambda$ as a
function of $\sigma$ at $\tau=100$, which is obtained by (i) directly
simulating the original model without external disturbances, (ii)
directly simulating the reduced phase model, (iii) calculation using
numerical phase maps assuming uniform phase PDF, and (iv) calculation
using numerical phase maps and the phase PDF approximated up to the
second order perturbation.
Since the phase map is always monotonically increasing in this range
of $\sigma$, the Lyapunov exponent $\lambda$ calculated assuming
uniform phase PDF (iii) is always non-positive and only becomes $0$
when $\sigma = 0$. The Lyapunov exponent $\lambda$ calculated using
approximate phase PDF (iv) is also always non-positive.
In this case, the correction to the uniform phase PDF is even smaller
than the previous Stuart-Landau case, hence the Lyapunov exponents
calculated by (iii) and by (iv) are almost indistinguishable. Of
course, they coincide with the results obtained by direct simulation
of the original model and the phase model.


\section{Summary}

We analyzed phase synchronization of general limit-cycle oscillators
subject to external impulses by reducing the dynamics of the
oscillator to a random phase map.
We proved that when the phase maps are strictly monotonic and the mean
inter-impulse interval of the input current is sufficiently large, the
Lyapunov exponent of the system always becomes negative, leading to
fluctuation-induced phase synchronization.
We also treated the case where the inter-impulse interval is finite
perturbatively for weak Poissonian impulses, and proved that the next
order correction to the upper bound of the Lyapunov exponent is also
zero, hence the fluctuation-induced phase synchronization persists
even if the phase distribution becomes slightly non-uniform.

Mathematically, the non-positivity of the Lyapunov exponent is a
general result of the concavity of the $\log$ function and the
monotonicity and periodicity of the phase map.
Therefore, this result is not restricted to specific oscillators, but
also holds generally for a wide variety of limit-cycle oscillators.
Examining the significance of our results in practical problems would
be an interesting topic.

Though we did not derive in this paper, we can reduce the phase model
driven by Poissonian impulses to an Ito-Langevin phase equation in the
limit of weak and frequent impulses when the net drift induced by the
external impulses vanishes. It yields
\begin{equation}
  \dot{\theta}(t) = \omega + {\bf Z}(\theta) \cdot \bm{\eta}(t),
\end{equation}
where $\bm{\eta}(t)$ is a $n$-dimensional Gaussian-white noise.
However, it can be shown that the phase disturbance is neutrally
stable for this Ito-Langevin phase model, namely, the Lyapunov
exponent is constantly zero~\cite{Teramae}, as a direct consequence of
the Ito stochastic integral~\cite{Gardiner}.
Therefore, if we take the Langevin limit within our phase model,
fluctuation-induced phase synchronization does not occur.
On the other hand, if the above Langevin phase equation is interpreted
in the Stratonovich sense, which {\it does not} come out of the
integration rule of the impulsive force we assumed in this paper, the
phase synchronization occurs~\cite{Teramae}.
Thus, slight difference in the treatment of the stochastic forcing
leads to physically distinct results. Detailed discussions on this
point, including the stochastic interpretation of impulsive forcing,
will be reported in the future.

\acknowledgments{We thank D. Tanaka, J. Teramae, T. Aoyagi, and S. Nii
  for useful discussions.}


\begin{thebibliography}{99}

\bibitem{Mainen} Z. F. Mainen and T. J. Sejnowski, Science {\bf 268}
  1503 (1995).

\bibitem{Steveninck} R. R. de Ruyter van Steveninck, G. D. Lewen,
  S. P. Strong, R. Koberle, W. Bialek, Science, {\bf 275} 1805 (1997).

\bibitem{Tsubo} Y. Tsubo, T. Kaneko. S. Shinomoto, Neural Networks
  {\bf 17} 165 (2004).
  
\bibitem{Jensen} R. V. Jensen, Phys. Rev. E. {\bf 58} R6907 (1998).

\bibitem{Kosmidis} E. K. Kosmidis and K. Pakdaman,
  J. Comput. Neurosci. {\bf 14} 5 (2003).

\bibitem{Pakdaman} K. Pakdaman, Neural Comput. {\bf 14} 781 (2002).
  
\bibitem{Gutkin} B. Gutkin, G. B. Ermentrout, and M. Rudolph,
  J. Comput. Neurosci. {\bf 15} 91 (2003).

\bibitem{Ritt} J. Ritt, Phys. Rev. E {\bf 68} 041915 (2003).

\bibitem{Casado}
  J. M. Casado and J. P. Baltan\'as, Int. J. Bif. and Chaos {\bf 14} 2061 (2004).

\bibitem{Teramae} J. N. Teramae and D. Tanaka, Phys. Rev. Lett. {\bf
    93} 204103 (2004).

\bibitem{Nagai} K. Nagai, H. Nakao, and Y. Tsubo, Phys. Rev. E {\bf
    71} 036217 (2005).

\bibitem{Goldobin} D. S. Goldobin and A. Pikovsky, Phys. Rev. E {\bf
    71} 045201(R) (2005); Physica A {\bf 351} 126 (2005).

\bibitem{Winfree} A. T. Winfree, {\it The Geometry of Biological Time}
  (Springer-Verlag, New York, 2001, 1980).
  
\bibitem{Kuramoto} Y. Kuramoto, {\it Chemical Oscillations, Waves, and
    Turbulence} (Springer-Verlag, Berlin, 1984).
  
\bibitem{Pikovsky2} A. Pikovsky, M. Rosenblum, and J. Kurths, {\it
    Synchronization} (Cambridge University Press, Cambridge, 2001).
  
\bibitem{Snyder} D. L. Snyder, {\it Random Point Processes} (John
  Wiley \& Sons, New York, 1975).
  
\bibitem{Hanggi} P. H\"anggi, Z. Physik B {\bf 31}, 407 (1978); Z.
  Physik B {\bf 36}, 271 (1980).

\bibitem{Lasota} A. Lasota and M. C. Mackey, {\it Probabilistic
    properties of deterministic systems} (Cambridge University Press,
  New York, 1985).
  
\bibitem{Pecora} L. M. Pecora and T. L. Carroll, Phys. Rev. A {\bf 44}
  2374 (1991).
  
\bibitem{Pikovsky} A. S. Pikovsky, Phys. Lett. A {\bf 165} 33 (1992).

\bibitem{Khoury} P. Khoury, M. A. Lieberman, and A. J. Lichtenberg,
  Phys. Rev. E. {\bf 54} 3377 (1996).

\bibitem{Toral} R. Toral, C.R. Mirasso, E. Hern\'andez-Garc\'ia,
  O. Piro, in {\it Unsolved Problems on Noise and Fluctuations},
  UPoN'99, D. Abbot and L. Kiss, eds. AIP Conference Proceedings 511,
  255 (2000).
  
\bibitem{Gardiner} C. W. Gardiner, {\it Handbook of Stochastic
    Methods} (Springer, Berlin, 1997).
  
\bibitem{Koch} C. Koch, {\it Biophysics of Computation} (Oxford
  University Press, Oxford, 1999).

\end{thebibliography}
\end{document}